\documentclass[12pt]{article}
																																																																																											\pdfoutput=1																																																																																			 
\usepackage{amsmath,amssymb}
\usepackage{graphicx}
\usepackage[pdftex]{hyperref}
\DeclareGraphicsExtensions{.eps,.bmp,.wmf,.jpeg,.pdf}
\numberwithin{equation}{section}
\def\be{\begin{equation}}
\def\ee{\end{equation}}

\def\bea{\begin{eqnarray}}
\def\eea{\end{eqnarray}}
\title{Exact solutions in a scalar-tensor model of dark energy}
\author{L. N. Granda\thanks{ngranda@univalle.edu.co}\, and \ E. Loaiza\thanks{edwin.loaiza@correounivalle.edu.co, Sede Buga}\\ {\small\it Departamento de Fisica, Universidad del Valle}\\{\small\it A.A. 25360, Cali, Colombia}}
\date{}
\begin{document}
\maketitle

\begin{abstract}
\noindent  We consider a model of scalar field with non minimal kinetic and Gauss Bonnet couplings as a source of dark energy. Based on asymptotic limits of the generalized Friedmann equation, we impose restrictions on the kinetic an Gauss-Bonnet couplings. This restrictions considerable simplify the equations, allowing for exact solutions unifying early time matter dominance with transitions to late time quintessence and phantom phases. The stability of the solutions in absence of matter has been studied.\\ 

\end{abstract}

\section{Introduction}
\noindent 
One of the most important challenges for the current theory of cosmology is the explanation of the late-time accelerated expansion of the universe. The cosmic acceleration has been supported by different observational data such as the supernovae type Ia \cite{riess}, \cite{perlmutter}, \cite{kowalski}, \cite{hicken}, cosmic microwave background anisotropy \cite{komatsu}, and large scale galaxy surveys \cite{percival}. The interpretation of astrophysical observations indicates that this accelerated expansion is due to some kind of  negative-pressure form of matter known as dark energy (DE). The combined analysis of cosmological observations also suggests that the universe is spatially flat, and  consists of about $\sim 1/3$ of dark matter, and $\sim 2/3$ of homogeneously distributed dark energy with negative pressure. 
The simplest candidate for dark energy is the cosmological constant, conventionally associated with the energy of the vacuum. However the cosmological constant presents the known severe energy scale problem \cite{peebles}, \cite{padmana1}. Among the different alternatives to explain the dark energy, the models involving a scalar field play an important role as they provide a dynamical behavior of the equation of state, which is  favored by astrophysical data (for a review see \cite{starovinski, copeland, sahnii, padmanabhan}).
A widely explored are scalar-tensor models, which contain a direct coupling of the scalar field to the curvature, providing in principle a mechanism to evade the coincidence problem and naturally allowing (in some cases) the crossing of the phantom barrier \cite{polarski}, \cite{peri}, \cite{maeda}. From a pure geometrical point of view, the modified gravity theories, which are generalizations of the general relativity, have been widely considered to describe the early-time inflation and late-time acceleration, without the introduction of any other dark component, and represent an important alternative to explain the dark energy (for review see \cite{sergei11}).\\
In the present work we consider a string and higher-dimensional gravity inspired scalar field model, with two types of couplings: kinetic coupling to curvatures and Gauss Bonnet (GB) coupling to the scalar field. The GB term is topologically invariant in four dimensions and does not contribute to the equations of motion. Nevertheless it affects cosmological dynamics when it is coupled to a dynamically evolving scalar field through arbitrary function of the field, giving rise to second order differential equations of motion, preserving the theory ghost free. Therefore, the coupled GB term seems as a natural generalization of the scalar field with non-minimal kinetic coupling to curvature \cite{granda,granda1}.\\
\noindent Some late time cosmological aspects of scalar field model with derivative couplings to curvature have been considered in \cite{sushkov}, \cite{granda,granda1,granda2}, \cite{saridakis}, \cite{gao}, \cite{granda3}. On the other hand, the GB invariant coupled to scalar field have been extensively studied. In \cite{sergei12}, \cite{koivisto} the GB correction was considered to study the dynamics of dark energy, where it was found that quintessence or phantom phase may occur in the late time universe. The evolution of perturbations in GB dark energy at large and small scales has been calculated in \cite{koivisto}, \cite{koivisto1}. Accelerating cosmologies with GB correction in four and higher dimensions have been discussed in \cite{tsujikawa}, \cite{leith}, \cite{maartens}. The modified GB theory applied to dark energy have been suggested in \cite{sergei14}, and different aspects of the modified GB model applied to late time acceleration, have been considered among others, in \cite{sergeio1}, \cite{sergeio2}, \cite{sergei15}, \cite{carter}, \cite{tretyakov}.\\
All these studies demonstrate that it is quite plausible that the scalar-tensor couplings predicted by a fundamental theory like string theory, may become important at current low-curvature universe (see \cite{sergeio3} and \cite{sergei11} for review).\\
The most general second-order ghost-free scalar-tensor Lagrangian with couplings to curvature, can be originated from toroidal compactification of $4+N$ dimensional Lagrangian of pure gravity, as shown in \cite{amendola}. In this compactification the scalar field plays the role of the overall size of the $N$-torus, and the couplings appear as exponentials of the scalar field. This general Lagrangian also appears in the next to leading order corrections in the $\alpha'$ expansion of the string theory \cite{tseytlin}, \cite{cartier}, \cite{meissner}.
In the present study we consider those terms in this ghost-free scalar-tensor Lagrangian that are coupled to the curvature, namely kinetic terms coupled to curvature and scalar field coupled to GB invariant, where we use more general couplings in order to find appropriate cosmological scenarios that satisfy the current astrophysical observations. Some late-time cosmological solutions have been studied in \cite{granda20}, and Big Rip and Little Rip solutions have been reconstructed in \cite{granda21}.
The main purpose of this work is to find exact solutions that are consistent with the large amount of existing data supporting the late-time accelerated expansion of the universe. Even in the Friedmann-Robertson-Walker background the cosmological equations are very difficult to integrate in an exact way, in part due to their non linearity and the variety of conditions that should be satisfied in order to cover the different evolutionary stages. Nevertheless, introducing restrictions on the interaction terms that are appropriate and consistent with early and late time cosmology, we can find suitable exact solutions describing early and late time asymptotic behavior. It was shown that the model with string (exponential) couplings in absence of matter, leads to power-law solution. 
In section II we introduce the model and give the general equations, which are then expanded on the FRW metric. In section III we consider two type of restrictions on the GB and Kinetic couplings in the case of scalar field dominance and found the respective solutions describing power-law, and more general viable late-time cosmologies. The issue of stability of these solutions is also addressed. In section IV we consider the matter contribution and study exact solutions unifying early time decelerated with late time accelerated expansion, including phantom phase. Concluding remarks are given in section V.
\section{Field Equations}
Let us start with the following action containing scalar field with kinetic couplings to curvature and the Gauss Bonnet coupling
\be\label{eq1}
\begin{aligned}
S=&\int d^{4}x\sqrt{-g}\Big[\frac{1}{16\pi G} R -\frac{1}{2}\partial_{\mu}\phi\partial^{\mu}\phi+F_1(\phi)G_{\mu\nu}\partial^{\mu}\phi\partial^{\nu}\phi- V(\phi)-F_2(\phi){\cal G}\Big]+S_m
\end{aligned}
\ee
\noindent where $G_{\mu\nu}=R_{\mu\nu}-\frac{1}{2}g_{\mu\nu}R$, ${\cal G}$ is the 4-dimensional GB invariant ${\cal G}=R^2-4R_{\mu\nu}R^{\mu\nu}+R_{\mu\nu\rho\sigma}R^{\mu\nu\rho\sigma}$ and $S_m$ is the matter action. The coupling $F_1(\phi)$ has dimension of $(length)^2$, and the coupling $F_2(\phi)$ is dimensionless. 
The GB coupling does not make contributions higher than second order (in the metric) to the equations of motion, and therefore does not introduce ghost terms into the theory. Hence, the equations derived from this action contain only second derivatives of the metric and the scalar field, avoiding problems with higher order derivatives \cite{capozziello1}.\\
Let us consider the spatially-flat Friedmann-Robertson-Walker (FRW) metric,
\be\label{eq2}
ds^2=-dt^2+a(t)^2\left(dr^2+r^2d\Omega^2\right)
\ee
Variation of the action (\ref{eq1}) with respect to the metric leads to the following equations (see \cite{granda2, granda3} for details)
\be\label{eq3}
H^2=\frac{\kappa^2}{3}\left(\rho_{DE}+\rho_m\right)
\ee
\be\label{eq4}
-2\dot{H}-3H^2=\kappa^2\left(p_{DE}+p_m\right)
\ee
where
\be\label{eq5}
\rho_{DE}=\frac{1}{2}\dot{\phi}^2+V(\phi)+9 H^2F_1(\phi)\dot{\phi}^2+24H^3\frac{dF_2}{d\phi}\dot{\phi}
\ee
and
\be\label{eq6}
\begin{aligned}
&p_{DE}=\kappa^2\Big[\frac{1}{2}\dot{\phi}^2-V(\phi)-\left(3H^2+2\dot{H}\right)F_1(\phi)\dot{\phi}^2\\&-2 H\left(2F_1(\phi)\dot{\phi}\ddot{\phi}+\frac{dF_1}{d\phi}\dot{\phi}^3\right)
-8H^2\frac{dF_2}{d\phi}\ddot{\phi}-8H^2\frac{d^2F_2}{d\phi^2}\dot{\phi}^2-16H\dot{H}\frac{dF_2}{d\phi}\dot{\phi}-16H^3\frac{dF_2}{d\phi}\dot{\phi}\Big]
\end{aligned}
\ee
where $\kappa^2=8\pi G$. In the present study we assume that the matter sector is modeled by an ideal fluid obeying the equation of state (EoS) $p_m=w\rho_m$ with constant parameter $w$ (mostly non relativistic matter with $p_m=0$), whose energy density satisfies the usual continuity equation $\dot{\rho_m}+3H\rho_m=0$ . The equation of motion for the scalar field takes the form
\be\label{eq7}
\begin{aligned}
&\ddot{\phi}+3H\dot{\phi}+\frac{dV}{d\phi}+3 H^2\left(2F_1(\phi)\ddot{\phi}+\frac{dF_1}{d\phi}\dot{\phi}^2\right)
+18 H^3F_1(\phi)\dot{\phi}+\\
&12 H\dot{H}F_1(\phi)\dot{\phi}+24\left(\dot{H}H^2+H^4\right)\frac{dF_2}{d\phi}=0
\end{aligned}
\ee
For generality we keep both $F_1(\phi)$ and $F_2(\phi)$ arbitrary, which allows to consider different viable cosmological scenarios containing quintessence and phantom phases (and  early time matter dominated phase, in the presence of matter). As we will see bellow, the exponential couplings (that are present in the next
to leading order corrections to string theory \cite{tseytlin}, \cite{cartier}, \cite{meissner}) appear from reasonable restrictions on the relative densities corresponding to kinetic and GB couplings. 
\section {The scalar field dominance}
We start studying solutions to Eqs. (\ref{eq3}-\ref{eq6}), in the important case when the scalar field potential and matter contribution are absent (i.e. $V=0$, $\rho_m=0$).  This leaves the two couplings $F_1(\phi)$ and $F_2(\phi)$ as the degrees of freedom that may characterize the late time cosmological dynamics. Considering the limit of dark energy dominance ($\rho_m\rightarrow 0$) and dark energy dominated by the kinetic coupling, we may neglect the free kinetic term and the GB coupling, leading from (\ref{eq3}) and (\ref{eq5}) to 
\be\label{eq8}
H^2\rightarrow 3F_1\dot{\phi}^2H^2
\ee
this limit suggests the following restriction for the kinetic coupling
\be\label{eq9}
F_1\dot{\phi}^2=k
\ee
where $k$ is constant. Note that if we divide the Eq. (\ref{eq3}) by $H^2$ then with the account of (\ref{eq5}) we may write Eq. (\ref{eq3}) in terms of the density parameters of DE and matter as
\be\label{eq10}
\Omega_{DE}+\Omega_m=\Omega_{\phi}+\Omega_{k}+\Omega_{GB}+\Omega_m=1
\ee
where 
\be\label{eq11}
\Omega_{\phi}=\frac{\rho_{\phi}}{3H^2},\,\,\,\,\Omega_{k}=3F_1\dot{\phi}^2,\,\,\,\,\,\,\Omega_{GB}=8H\frac{dF_2}{d\phi}\dot{\phi}=8H\frac{dF_2}{dt},\,\,\,\Omega_m=\frac{\rho_m}{3H^2}
\ee
where we have set $\kappa^2=1$ and $\rho_{\phi}=\dot{\phi}^2/2+V$, $\Omega_{k}$ and $\Omega_{GB}$ are the density parameters associated with the kinetic and GB couplings respectively. 
Considering the limit of DE dominated by the GB coupling (i.e. $\dot{\phi}\rightarrow 0$) and , then from (\ref{eq3}) and (\ref{eq5}) follows
\be\label{eq12}
H^2\rightarrow 8H^3\frac{dF_2}{d\phi}\dot{\phi}
\ee
which motivates the restriction on the GB coupling
\be\label{eq13}
\frac{dF_2}{dt}=\frac{g}{H(t)}
\ee
Then, considering the definitions in (\ref{eq11}), we may interpret the restrictions on the kinetic and GB couplings as the condition that the density parameters  $\Omega_{k}$ and $\Omega_{GB}$ become constants.
\be\label{eq13a}
\Omega_k=3k,\,\,\,\,\,\, \Omega_{GB}=8g
\ee
where $k$ and $g$ are constants. Having fixed the couplings $F_1$ and $F_2$ we can solve the Eqs. (\ref{eq3})-(\ref{eq7}).\\
An important consequence of the couplings of the form (\ref{eq9}) and (\ref{eq13}) comes from the generalized Friedmann equation (\ref{eq3}, \ref{eq5}). From this equation and making $V=0, \rho_m=0$, follows 
\be\label{eq14}
H^2=\frac{1}{3}\left(\frac{\dot{\phi}^2}{2}+9k H^2+24g H^2\right)
\ee
which imposes the behavior of the kinetic term 
\be\label{eq15}
\dot{\phi}^2=\lambda^2H^2
\ee
where $\lambda^2=6\left(1-3k-8g\right)$, which should be positive, i.e. $1-3k-8g>0$. This also indicates that the kinetic term scales in the same way as $\rho_{DE}$. Note that if we take $1-3k-8g=1$, then the important limit $\dot{\phi}^2=6H^2$ is obtained, which in absence of kinetic and GB couplings would correspond to purely kinetic model describing ``stiff'' matter with equation of state $w=1$. But the scalar field should satisfy additionally the equation of motion (\ref{eq7}) which depends on the couplings. Multiplying the Eq. (\ref{eq7}) by $\dot{\phi}$ we can reduce it to a first order equation for the function $\dot{\phi}^2$ as follows (making $V=0$)
\be\label{eq16}
\frac{1}{2}\frac{d(\dot{\phi}^2)}{dt}+3H\dot{\phi}^2+3H^2\frac{d}{dt}(F_1\dot{\phi}^2)
+18 H^3F_1\dot{\phi}^2+12 H\dot{H}F_1\dot{\phi}^2+24\left(\dot{H}H^2+H^4\right)\frac{dF_2}{dt}=0
\ee
Replacing $F_1\dot{\phi}^2$ from (\ref{eq9}) and $dF_2/dt$ from (\ref{eq13}) we find
\be\label{eq17}
\frac{1}{2}\frac{d(\dot{\phi}^2)}{dt}+3H\dot{\phi}^2+18kH^3+12kH\dot{H}+24gH\left(\dot{H}+H^2\right)=0
\ee
and taking into account the Eq. (\ref{eq15}), gives
\be\label{eq18}
\left(1-k-4g\right)\dot{H}+\left(3-6k-20g\right)H^2=0
\ee
this equation has power-law solution 
\be\label{eq19}
H=\frac{p}{t},\,\,\,\,\,\,\, p=\frac{1-k-4g}{3-6k-20g}
\ee
leading to accelerated expansion provided $p>1$. If we take into account the integration constant, then this solution could describe phantom behavior with future Big Rip singularity. Note that by setting $3-6k-20g=0$, leads to the de Sitter solution $H=const.$. Using (\ref{eq19}) in Eq. (\ref{eq15}) we find the scalar field as
\be\label{eq20}
\phi=\lambda p\ln t
\ee
and the kinetic and GB coupling from (\ref{eq9}) and (\ref{eq13}) respectively as
\be\label{eq21}
F_1=\frac{k}{\lambda^2p^2}e^{2\phi/(\lambda p)},\,\,\,\,\, F_2=\frac{g}{2p}e^{2\phi/(\lambda p)}
\ee
where we used (\ref{eq20}) for $t$. Note that this exponential behavior is exactly what is expected from string theory for the kinetic and GB couplings \cite{tseytlin}, \cite{cartier} (appropriately rescaling the scalar field).
An important result from Eq. (\ref{eq18}) is that it can also be satisfied when the two constant coefficients are zero. This leads to two linear equations for $k$ and $g$ with solution: $k=-2, g=3/4$.
For this concrete values the Eq. (\ref{eq17}) ((\ref{eq18}))satisfies automatically for any $H(t)$. This fact can be used to reconstruct the model for any given cosmological evolution encoded in $H(t)$. In any case if the Hubble parameter differs from power-law, the reconstructed couplings $F_1$ and $F_2$ would be different from simple exponentials. As we have seen the restrictions (\ref{eq13a}) are reasonable approximations for late-time cosmology.\\
We also can find exact solutions for the equations ((\ref{eq3}-\ref{eq6})) by considering the restrictions
\be\label{eq22a}
\Omega_k=c_k a^{-\alpha},\,\,\,\,\,\,\, \Omega_{GB}=c_g a^{-\alpha}
\ee
where $c_k$ and $c_g$ are numerical constants. This restriction for the GB coupling has been considered in \cite{neupane}. Introducing the e-folding variable $N=\ln a$ the equations 
(\ref{eq3}) and (\ref{eq7}) take the form
\be\label{eq22b}
H^2=\frac{1}{3}\left[\frac{1}{2}H^2\theta+V+9H^4 F_1\theta+24H^4\frac{dF_2}{dN}\right]
\ee
and
\be\label{eq22c}
\begin{aligned}
&\frac{1}{2}\frac{d}{dN}(H^2\theta)+3H^2\theta+\frac{dV}{dN}+9 H^2(F_1\theta)\frac{dH^2}{dN}+3 H^4\frac{d}{dN}(F_1\theta)\\
&+18 H^4(F_1\theta)+12H^2\frac{dH^2}{dN}\frac{dF_2}{dN}+24H^4\frac{dF_2}{dN}=0
\end{aligned}
\ee
where we have used $d/dt=H d/dN$. We have multiplied the Eq. (\ref{eq17}) by $\dot{\phi}$ and represented $(d\phi/dN)^2=\theta(N)$. Using the restrictions (\ref{eq22a}) and additionally considering that the scalar field evolves as $\theta=const.=\lambda^2$ (i.e. $\phi=\lambda N +\phi_0$), then from (\ref{eq22b}) follows
\be\label{eq22d}
V=H^2\left(3-\lambda^2/2-3(c_k +c_g)e^{-\alpha N}\right)
\ee
Setting $\lambda^2=6$ to simplify the expression for the potential (i.e. $V=-3(c_k +c_g)H^2e^{-\alpha N}$) and replacing in the Eq. (\ref{eq22c}), then the equation for $H^2$ becomes
\be\label{eq22e}
3\frac{dH^2}{dN}+18H^2-(c_k+\frac{3}{2}c_g)e^{-\alpha N}\frac{dH^2}{dN}+\left(2(3+\alpha)c_k+3(1+\alpha)c_g\right)e^{-\alpha N}H^2=0
\ee
Solving this equation one finds
\be\label{eq22f}
H^2=C e^{\eta_1 N}\left[3c_g+2c_k-6 e^{\alpha N}\right]^{\eta_2}
\ee
where $C$ is the integration constant and 
\be\label{eq22g}
\eta_1=2\left(\frac{3c_g+6c_k}{3c_g+2c_k}+\alpha\right),\,\,\,\,\eta_2=-2-\frac{24(c_g+c_k)}{(3c_g+2c_k)\alpha}
\ee
From (\ref{eq22a}) and (\ref{eq22f}) one finds the expressions for the kinetic and GB couplings
\be\label{eq22f1}
F_1=\frac{c_k}{18C}e^{-(\alpha+\eta_1)N}\left[3c_g+2c_k-6 e^{\alpha N}\right]^{-\eta_2}
\ee
and
\be\label{eq22f2}
F_2=-\frac{c_g e^{-(\alpha+\eta_1)N}}{8C(\alpha+\eta_1)(3c_g+2c_k)^{\eta_2}}\hspace{0.2cm} _{2}F_1\left[-\frac{\alpha+\eta_1}{\alpha},\eta_2,-\frac{\eta_1}{\alpha},\frac{6 e^{\alpha N}}{3c_g+2c_k}\right]
\ee
Note that the value $\eta_2=0$ leads to the solution known as ``stiff'' matter $H^2\propto e^{-6 N}\sim a^{-6}$. Interesting late-time cosmological solutions can be found by 
setting the power $\eta_2=1$, or $2$. Solving the condition $\eta_2=1$ with respect to $\alpha$ one finds
\be\label{eq22h}
H^2=C e^{-2\frac{5c_g+2c_k}{3c_g+2c_k}N}\left(3c_g+2c_k-6 e^{-\frac{8(c_g+c_k)}{3c_g+2c_k} N}\right)
\ee
Note that all found solutions may be written in terms of the scalar field by replacing $N=\phi/\sqrt{6}$. In order for this solution to make sense, the integration constant $C$ should be negative and $3c_g+2c_k<0$. Combining this last inequality with $c_g+c_k<0$ (that follows from the positivity of the potential (\ref{eq22d}) for $\lambda^2=6$) leads to the restrictions ($c_g\le 0$ and $c_k\le -c_g$) or ($c_g>0$ and $c_k<-3c_g/2$).  Under this conditions, the argument of the second exponent in (\ref{eq22h}) becomes negative and the sign in the argument of the first exponent depends on the particular values of $c_g$ and $c_k$. Therefore the solution (\ref{eq22h}) is able to explain the late-time cosmic acceleration. In terms of the scalar field the potential becomes
\be\label{eq22i}
V=18C(c_g+c_k) e^{\beta_1 \phi}-3C(c_g+c_k)(3c_g+2c_k) e^{\beta_2 \phi}
\ee
where
\be\nonumber
\beta_1=-2\frac{5c_g+2c_k}{\sqrt{6}(3c_g+2c_k)},\,\,\,\,\, \beta_2=-\frac{2(c_g-2c_k)}{\sqrt{6}(3c_g+2c_k)}
\ee
Considering $\eta_2=2$, and solving this condition with respect to $\alpha$ gives
\be\label{eq22j}
H^2=C e^{-\frac{\sqrt{6}c_g}{3c_g+2c_k}\phi}\left(3c_g+2c_k-6 e^{-\frac{\sqrt{6}(c_g+c_k)}{3c_g+2c_k} \phi}\right)^2
\ee
where $C>0$ and the only restriction on the constants $c_g$ and $c_k$ is $c_g+c_k<0$, to keep the positivity of the potential. Therefore the present solution is able to explain late-time cosmic acceleration, including super-acceleration ($w_{eff}<-1$) if one of the arguments in the exponentials in (\ref{eq22j}) is positive. The scalar potential takes the form
\be\label{eq22k}
V=V_0 e^{\gamma_1\phi}\left(3c_g+2c_k-6 e^{\gamma_2 \phi}\right)^2
\ee
where $V_0=-3C(c_g+c_k)$, $\gamma_1=\frac{\sqrt{6}c_k}{3c_g+2c_k}$, $\gamma_2=-\frac{\sqrt{6}(c_g+c_k)}{3c_g+2c_k}$. 
\subsection*{Stability of solutions}
To board the issue of stability of the solutions corresponding to the restrictions (\ref{eq13a}) and (\ref{eq22a}) we will introduce the autonomous system for the Eqs. (\ref{eq3})-(\ref{eq7}) in absence of matter. We used this approach because of the nature of the restrictions we have considered, that have direct connection with the dynamical variables defined for this model. The stability analysis based on perturbations, for inflationary solutions in more general second-order scalar tensor theory, has been performed in \cite{kobayashi}. Let's consider the following dynamical variables
\be\label{eqdy1}
\begin{aligned}
&x=\frac{\kappa\dot{\phi}}{\sqrt{2}H}, \,\,\, y=\frac{\kappa^2 V}{H^2},\,\,\, k=3\kappa^2F_1\dot{\phi}^2, \\
& g=8\kappa^2 H\dot{\phi}\frac{dF_2}{d\phi},\,\,\, \,\,\, \epsilon=\frac{\dot{H}}{H^2}
\end{aligned}
\ee
The terms with derivatives in (\ref{eq3})-(\ref{eq7}) transform under the change of variable as
\be\label{eqdy2}
\begin{aligned}
\ddot{\phi}&=\frac{\sqrt{2} H^2 (x \epsilon +x')}{\kappa }\\
\frac{dV}{d\phi}&=\frac{H^2 (2 y \epsilon +y')}{\sqrt{2} \kappa  x}\\
\frac{d^2F_2}{d\phi^2}&=\frac{x g'-g (2 x \epsilon +x')}{16 H^2 x^3}\\
\frac{dF_1}{d\phi}&=\frac{x k'-\kappa  k (x \epsilon +x')}{3 \sqrt{2} H^2 x^4}
\end{aligned}
\ee
where ``prime'' denotes the derivative w.r.t the e-folding variable $N$. In the subsequent analysis we will set $\kappa^2=1$.
In terms of the variables (\ref{eqdy1}) and using (\ref{eqdy2}), the Eqs. (\ref{eq3})-(\ref{eq7}) can be transformed into the following first-order autonomous system
\be\label{eqdy3}
\begin{aligned}
x^2+y+3k+3g-3=0&=0\\
2x x'+2(3+\epsilon)x^2+y'+2\epsilon y+k'+2(3+2\epsilon)k+3(1+\epsilon)g&=0\\
2\epsilon +3+x^2-y-\frac{2}{3}k'-\frac{1}{3}(3+2\epsilon)k-g'-(2+\epsilon)g&=0
\end{aligned}
\ee
We should consider that the solution (\ref{eq19})-(\ref{eq21}) was obtained without potential (i.e. $V=0$ in (\ref{eqdy1}) and therefore there is not variable $y$ associated with the potential. Additionally we have to take into account the constraints on the dynamical variables $k$ and $g$ (\ref{eq13a}), that lead to the solution (\ref{eq19})-(\ref{eq21}). All this amounts to complement the dynamical system (\ref{eqdy3})  with the following equations
\be\label{eqdy3a1}
y=0, \;\;\;\, g'=0,\;\;\;\;\, k'=0
\ee
And therefore the system reduces to the following one-dimensional dynamical system for $x$
\be\label{eqdy3a}
3 g+3 k+x^2-3=0
\ee
\be\label{eqdy3b}
3 g (\epsilon +1)+2 \epsilon (2 k+x^2)+6 k+2 x (3 x+x')=0
\ee
where
\be\label{eqdy3c}
\epsilon=-\frac{3 \left(2 g+k-x^2-3\right)}{3 g+2 k-6}
\ee
Note that the constants $k$ and $g$ are not arbitrary, as they should satisfy the restriction (\ref{eqdy3a}). From (\ref{eqdy3b}) and (\ref{eqdy3c}) follows
\be\label{eqdy6}
x'=f(x)=\frac{3 \left(g-2 x^2\right) \left(3 (g+k-1)+x^2\right)}{2 x (3 g+2 k-6)}\\
\ee
The critical points of  (\ref{eqdy6}) are: $x_{1,2}=\pm\sqrt{g/2}$ and $x_{3,4}= \pm \sqrt{3(1-g-k)}$. It is easy to check that the solution (\ref{eq19}), (\ref{eq20}) is a critical point provided $\lambda=\sqrt{2}x_{1,2}$ or $\lambda=\sqrt{2}x_{3,4}$. Taking the derivative of (\ref{eqdy6}) w.r.t. $x$ and evaluating at the critical points, we find the eigenvalue equations
\begin{enumerate}
\item $\lambda_1=\frac{df}{dx}|_{x=x_{1,2}}=-\frac{3 (7 g+6 k-6)}{3 g+2 k-6}$. The stability of the fixed point $x_{1,2}$ demands that $\lambda_1<0$, which leads to 
one of the following conditions: ($g>6$, $k>\frac{1}{6} (6-7 g)$), ($g\leq6$, $k>\frac{1}{2}(6-3g)$), ($g\leq6,k<\frac{1}{6}(6-7g)$) or ($g>6,k<\frac{1}{2}(6-3g)$). If one of this conditions is satisfied, then the fixed point $x_{1,2}$ is an attractor.
\item  $\lambda_2=\frac{df}{dx}|_{x=x_{3,3}}=\frac{3 (7 g+6 k-6)}{3 g+2 k-6}$. For $\lambda_2<0$ this fixed point is an attractor, which leads to the conditions:
($g<6$, $\frac{1}{6}(6-7g)<k<\frac{1}{2}(6-3g)$) or ($g>6$, $\frac{1}{2}(6-3g)<k<\frac{1}{6}(6-7g)$).
\end{enumerate}
Taking into account that the solution is a fixed point $x=x_0=\lambda/\sqrt{2}$, then $\lambda$ takes the following values: $\lambda=\pm\sqrt{g}$ for the fixed points $x_{1,2}$ and $\lambda=\pm\sqrt{6(1-k-g)}$ for $x_{3,4}$. Using this results in the Eq. (\ref{eqdy3a}), then the above inequalities translate into the following conditions on $g$ to achieve the stability of the solution (\ref{eq19})-(\ref{eq21}):
\be\label{eqdy7}
\lambda_1<0: \;\;\;\; 0<g<12
\ee
\be
\lambda_2<0:\;\;\;\; g<0 \,\,\,\, or \,\,\,\,\, g>12
\ee
For the second solution we have used the restrictions (\ref{eq22a}) on the density parameters for the kinetic and Gauss-Bonnet couplings as defined in (\ref{eq11}). Note that these density parameters coincide with the ones defined as the corresponding dynamical variables in (\ref{eqdy1}) (i.e. $k\equiv\Omega_k$ and $g\equiv\Omega_g$, setting $\kappa^2=1$). In addition, to obtain the solution (\ref{eq22f}) we have introduced the ansatz:
\be\label{eqdy8}
\theta=\phi'^2=\lambda^2
\ee
where we have limited to the case $\lambda^2=6$, which gives the potential of the form
\be\label{eqdy9}
V=V_0 H^2 e^{-\alpha N},\,\,\,\,\,\,\, V_0=-3(c_k+c_g)
\ee
Form Eq. (\ref{eqdy8})  follows that
\be\label{eqdy10}
\phi=\phi_0+\sqrt{6} N
\ee
And therefore the variable $x$ becomes constant as follows from the definition $x=\frac{\dot{\phi}}{\sqrt{2}H}=\frac{\phi'}{\sqrt{2}}=\sqrt{3}=x_0$. This leads to $x'=0$. 
Setting $\phi_0=0$, we can re express the variable N in terms of $\phi$, and the potential takes the form: $V=V_0H^2 e^{-\alpha\phi/\sqrt{6}}$, which according to (\ref{eqdy1}) gives $y=V_0 e^{-\alpha\phi/\sqrt{6}}$, and from (\ref{eq22a}): $k=c_k e^{-\alpha\phi/\sqrt{6}}$, $g=c_g e^{-\alpha\phi/\sqrt{6}}$. Using this results, the dynamical system reduces to the equations
\be\label{eqdy11}
y'=-\frac{\alpha}{\sqrt{3}}x y\,\,\,\, g'=-\frac{\alpha}{\sqrt{3}}x g,\,\,\,\,\,\, k'=-\frac{\alpha}{\sqrt{3}}x k
\ee
From these equations follows that the only critical point is $(0,0,0)$ and the eigenvalue matrix becomes diagonal with all the eigenvalues equal to 
\be\label{eqdy12}
\lambda_1=\lambda_2=\lambda_3=-\frac{\alpha}{\sqrt{3}}x_0=-\alpha
\ee
from which follows that the point $(0,0,0)$ is a stable critical point (attractor) if $\alpha>0$. This makes sense if we see that at $t\rightarrow\infty$ (or $a\rightarrow\infty$) (as follows from the solution (\ref{eq22f}) for the discussed restrictions on $c_k$ and $c_g$), then $g\rightarrow 0$, $k\rightarrow 0$ and $y\rightarrow 0$ for positive $\alpha$, as follows from (\ref{eq22a}) and (\ref{eqdy9}). 
\subsection*{Stability under tensor and scalar perturbations}
The present model is a particular case of the generalized Galileon theory \cite{kobayashi}, \cite{felice}, \cite{felice1}, \cite{felice2}. The conditions to avoid ghost and gradient instabilities in the generalized Galileon theory have been presented in \cite{kobayashi}. Here we use the stability analysis presented in \cite{kobayashi} applied to the model (\ref{eq1}) as special case in absence of matter. 
The generalized Galileon model may be written as \cite{kobayashi} (setting $\kappa^2=1$)
\be\label{eqggs}
\begin{aligned}
S=&\int d^4x\sqrt{-g}\Big[K(\phi,X)-G_3(\phi,X)\Box\phi+G_4(\phi,X)R+\frac{\partial G_4}{\partial X}\left((\Box\phi)^2-(\nabla_{\mu}\nabla_{\nu}\phi)^2 \right)\\
&+G_5(\phi,X)G_{\mu\nu}\nabla{\mu}\nabla^{\nu}\phi-\frac{1}{6}\frac{\partial G_5}{\partial X}\left[(\Box\phi)^3-3(\Box\phi)(\nabla_{\mu}\nabla_{\nu}\phi)^2+2(\nabla_{\mu}\nabla_{\nu}\phi)^3\right]\Big]
\end{aligned}
\ee
where $X=-\nabla_{\mu}\phi\nabla^{\mu}\phi/2$ and $(\nabla_{\mu}\nabla_{\nu}\phi)^3=(\nabla_{\mu}\nabla_{\nu}\phi)(\nabla^{\nu}\nabla^{\sigma}\phi)(\nabla_{\sigma}\nabla^{\mu}\phi)$ . To obtain the model (\ref{eq1}) (with $S_m=0$) we use the following correspondence:
\be\label{eqggs1}
\begin{aligned}
&K(\phi,X)=-V(\phi)+X-8\frac{d^4F_2}{d\phi^4} X^2(3-\ln X),\,\,\,\, G_3=-4\frac{d^3F_2}{d\phi^3}X(7-3\ln X)\\
&G_4=\frac{1}{2}-4\frac{d^2F_2}{d\phi^2}X(2-\ln X),\,\,\,\, G_5=-\frac{\phi F_1(\phi)}{1+\phi\frac{d(\ln F_1)}{d\phi}}+4\frac{dF_2}{d\phi}\ln X
\end{aligned}
\ee
where the first term in the expression for $G_5$ gives (up to total derivative) the non-minimal kinetic coupling as appears in the third term in (\ref{eq1}) (note that $F_1=1$ leads to $G_5=-\phi$), and all the terms depending on $F_2$ in (\ref{eqggs1}) reproduce the GB coupling in (\ref{eq1}).\\ 

\noindent {\bf \it Tensor perturbations}\\
The quadratic action for tensor perturbations $h_{ij}$ is given by (see \cite{kobayashi})
\be\label{eqggs2}
\delta^2S_T=\frac{1}{8}\int dt d^3x a^3\left({\cal G}_T\dot{h}_{ij}^2-\frac{{\cal F}_T}{a^2}(\vec{\nabla} h_{ij})^2\right)
\ee
where
\be\label{eqggs3}
\begin{aligned}
&{\cal G}_T=2\left[G_4-2X\frac{\partial G_4}{\partial X}-X\left(H\dot{\phi}\frac{\partial G_5}{\partial X}-\frac{\partial G_5}{\partial \phi}\right)\right],\\
&{\cal F}_T=2\left[G_4-X\left(\ddot{\phi}\frac{\partial G_5}{\partial X}+\frac{\partial G_5}{\partial \phi}\right)\right]
\end{aligned}
\ee
from (\ref{eqggs2}) follows that the conditions to avoid ghost and gradient instabilities under tensor perturbations reduce to 
\be\label{eqggs4}
{\cal G}_T>0,\;\;\;\, {\cal F}_T>0
\ee
Replacing the solutions (\ref{eq19})-(\ref{eq21}) in (\ref{eqggs3}), taking into account the definitions (\ref{eqggs1}), we can analyze the conditions (\ref{eqggs4}). The condition  ${\cal G}_T>0$ takes the from
\be\label{eqggs4a}
{\cal G}_T=1-8g-\frac{k(\lambda^2 p^2+2\lambda p\phi+4\phi^2)}{(\lambda p+2\phi)^2}>0
\ee
Then the tensor ghost is absent if the above inequality is satisfied, or if $g<0$ and $k<0$ (provided $\phi=\lambda p \ln t>0$). Note that at large times ${\cal G}_T$ behaves as
\be\label{eqggs4b}
{\cal G}_T=1-8g-k
\ee
which is also the de Sitter limit that takes place at $p\rightarrow\infty$ (in this limit the EoS parameter $w=-1+2/(3p)$ takes the value $w=-1$). So the condition $k<1-8g$ guarantees the absence of ghosts instabilities at large times and in the de Sitter limit. The condition for ${\cal F}_T$ is given by
\be\label{eqggs4c}
{\cal F}_T=1-\frac{8g}{p}+\frac{k(\lambda^2 p^2+2\lambda p\phi+4\phi^2)}{(\lambda p+2\phi)^2}>0
\ee
at large times and in the de Sitter limit ${\cal F}_T$ behaves respectively as 
\be\label{eqggs4d}
{\cal F}_T|_{t\rightarrow\infty}=1-\frac{8g}{p}+k,\;\;\;\, {\cal F}_T|_{p\rightarrow\infty}=1+k
\ee
Combining with the previous conditions, the avoidance of ghost and gradient instabilities under tensor perturbations impose the following restrictions at large times:
\be\label{eqggs4e}
p>0,\;\;\;\, g<\frac{p}{4+4p},\;\;\;\, \frac{8g-p}{p}<k<1-8g 
\ee
and at the de Sitter limit we find the following restrictions
\be\label{eqggs4f}
g<\frac{1}{4},\;\;\;\;\;\, -1<k<1-8g
\ee
Anther interesting result is obtained in the case when we neglect the kinetic coupling. By setting $k=0$ in (\ref{eqggs4a}) and (\ref{eqggs4c}) it follows that ${\cal G}_T$ and ${\cal F}_T$ become constants equal to
\be\label{eqggs4g}
{\cal G}_T=1-8g,\,\,\,\,\, {\cal F}_T=1-\frac{8g}{p}
\ee
Thus in the particular case without kinetic coupling, the model is free of ghost and gradient instabilities  (independently of time) under tensor perturbations, for the power-law evolution, provided $g$ satisfies the restrictions
\be\label{eqggs4h}
0<p<1,\,\,\,\, g<p/8,\,\,\, \text{or} ,\,\,\, p>1, \,\,\,\, g<1/8
\ee
Let's consider the second solution given by Eqs. (\ref{eq22d}) (with $\lambda^2=6$), (\ref{eq22f})-(\ref{eq22f2}) in the $N$ variable. To evaluate the conditions (\ref{eqggs4}) for this solution, we use $d/dt=H d/dN$ in Eqs. (\ref{eqggs1}) and (\ref{eqggs3}). In this case the expressions for ${\cal G}_T$ and ${\cal F}_T$
are more involved, but assuming $\alpha>0$ we have found that at large times  (future infinity) ${\cal G}_T\rightarrow 1$ and ${\cal F}_T\rightarrow 1$ (independently of $c_g$ and $c_k$), which indicates that we can expect stability under tensor perturbations at future times. On the other hand, if we consider $\alpha<0$ then the stability conditions take place at early times. Nevertheless, these appreciations correspond to asymptotic behavior. To give a numerical example we analyze the specific viable model of DE  (\ref{eq22j}), (\ref{eq22k}) corresponding to $\eta_2=2$. To simplify the numerical analysis we use the units $\kappa^2=1, H_0^2=1$. As initial condition we will assume that the current value of the EoS is $w=-1$ and the constants $c_g,c_k$ and $C$ are subject to the flatness condition (here the constant $C$ is given in units of $H_0^2$). Additionally we can restrict the constants $c_g$ and $c_k$ in such way that the ``stiff'' matter term ($\propto a^{-6}$) in the expression for $H$ disappears. This gives the relation $c_k=-\frac{3 (g^2- 8g)}{2 (g-6)}$. The numerical analysis for the specific values of $c_g=10^{-12}$ and $c_k=-2\times 10^{-12}$ is resumed in Fig.1. For these values the DE model describes with high accuracy a de Sitter universe, with EoS practically constant, $w=-1$.
\begin{center}
\includegraphics [scale=0.7]{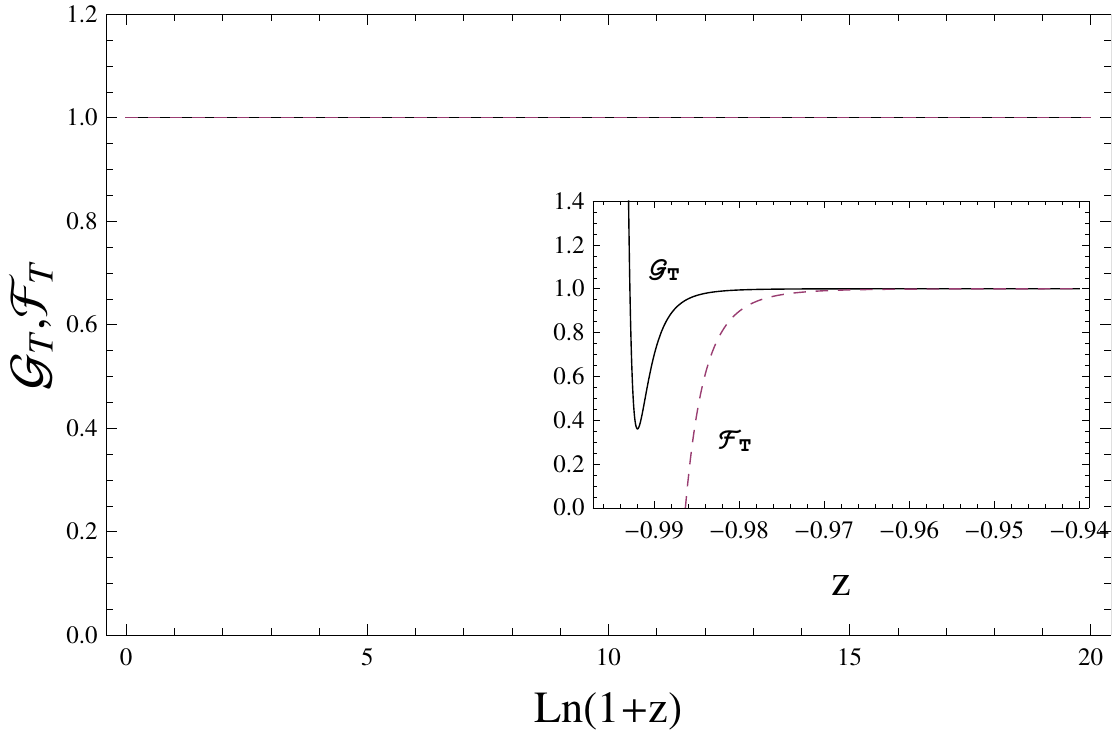}
\end{center} 
\begin{center}
{Fig. 1 \it ${\cal G}_T$ and ${\cal F}_T$ versus $\ln(1+z)$, for $c_g=10^{-12}$ and $c_k=-2\times 10^{-12}$. Note that ${\cal G}_T$ and ${\cal F}_T$ are practically constants equal to $1$, during all cosmological epochs. The internal graphic shows the future behavior of ${\cal G}_T$ and ${\cal F}_T$ for the redshift $z$ in the region $(-0.99,-0.94)$. Note that starting form $z\sim -0.97$ the values of ${\cal G}_T$, ${\cal F}_T$ begin to differ from $1$. This solution is stable under tensor perturbations for a wide range of redshifts} 
\end{center}
So we see that the solution is stable under tensor perturbations during different cosmological epochs, including late times and future evolution, and starting at $z\sim -0.98$, ${\cal F}_T$ tends to negative values, signaling instabilities at far future when the Eos tends to $w|_{z\rightarrow -1}=-3$.\\

\noindent {\bf \it Scalar perturbations}\\
A similar expression for the quadratic action for scalar perturbations $\zeta$ is given by 
\be\label{eqggs5}
\delta^2S_S=\int dt d^3x a^3\left({\cal G}_S\dot{\zeta}^2-\frac{{\cal F}_S}{a^2}(\vec{\nabla}\zeta)^2\right)
\ee
where
\be\label{eqggs6}
\begin{aligned}
&{\cal G}_S=\frac{\Sigma}{\Theta^2}{\cal G}_T^2+3{\cal G}_T\\
&{\cal F}_S=\frac{d}{dt}\left(\frac{{\cal G}_T^2}{\Theta}\right)+H\frac{{\cal G}_T^2}{\Theta}-{\cal F}_T
\end{aligned}
\ee
with $\Sigma$ and $\Theta$ given by 
\be\label{eqggs7}
\begin{aligned}
\Sigma=&X\frac{\partial K}{\partial X}+2X^2\frac{\partial^2K}{\partial X^2}+6H\dot{\phi}X\left(2\frac{\partial G_3}{\partial X}+X\frac{\partial^2G_3}{\partial X^2}\right)
-2X\left(\frac{\partial G_3}{\partial \phi}+X\frac{\partial^2 G_3}{\partial \phi\partial X}\right)\\&-6H^2G_4+6\Big[H^2X\left(7\frac{\partial G_4}{\partial X}+16X\frac{\partial^2 G_4}{\partial X^2}+4X^2\frac{\partial^3 G_4}{\partial X^3}\right)\\
&-H\dot{\phi}\left(\frac{\partial G_4}{\partial \phi}+5X\frac{\partial^2 G_4}{\partial \phi\partial X}+2X^2\frac{\partial^3 G_4}{\partial \phi\partial X^2}\right)\Big]+H^3\dot{\phi}X\Big(30\frac{\partial G_5}{\partial X}+26X\frac{\partial^2 G_5}{\partial X^2}\\
&+4X^2\frac{\partial^3 G_5}{\partial X^3}\Big)-6H^2X\left(6\frac{\partial G_5}{\partial \phi}+9X\frac{\partial^2 G_5}{\partial \phi\partial X}+2X^2\frac{\partial^3 G_5}{\partial \phi\partial X^2}\right)
\end{aligned}
\ee
\be\label{eqggs8}
\begin{aligned}
\Theta=&-\dot{\phi}X\frac{\partial G_3}{\partial X}+2H\left(G_4-4X\frac{\partial G_4}{\partial X}-4X^2\frac{\partial^2 G_4}{\partial X^2}\right)+\dot{\phi}\frac{\partial G_4}{\partial \phi}\\
&+2X\dot{\phi}\frac{\partial^2 G_4}{\partial \phi\partial X}-H^2\dot{\phi}\left(5X\frac{\partial G_5}{\partial X}+2X^2\frac{\partial^2 G_5}{\partial X^2}\right)+2HX\left(3\frac{\partial G_5}{\partial \phi}+2X\frac{\partial^2 G_5}{\partial \phi\partial X}\right)
\end{aligned}
\ee
The ghost and gradient instabilities are absent provided 
\be\label{eqggs9}
{\cal G}_S>0,\;\;\;\;\, {\cal F}_S>0
\ee
In the case of the solution (\ref{eq19})-(\ref{eq21}), and using (\ref{eqggs1}) and (\ref{eqggs6})-(\ref{eqggs8}), we find
\be\label{eqggs10}
\begin{aligned}
{\cal G}_S=& \Big[3 \Big(4 (8 g + k - 1) \ln^2(t) + 2 (16 g + k - 2) \ln(t)\\
& + 8 g + k - 
       1\Big) \Big(16 (64 g^2 + 8 g (4 k - 1) - 6 k^2 + 3 k - 1) \ln^4(t) + 
     8 (256 g^2 + 8 g (9 k - 4)\\
		&- 15 k^2 + 12 k - 4) \ln^3(t) + 
     4 (384 g^2 + 80 g k - 48 g - 21 k^2 + 18 k - 6) \ln^2(t) \\
		&+ 
     2 (256 g^2 + 8 g (9 k - 4) - 15 k^2 + 12 k - 4) \ln(t) + 
     64 g^2 + 8 g (4 k - 1) - 6 k^2 + 3 k - 1\Big)\Big]\\
		&/\left[(2 \ln(t) + 
     1)^2 (4 (3 k - 1) \ln^2(t) + (6 k - 4) \ln(t) + 3 k - 1)^2\right]
		\end{aligned}
		\ee
		
		and 
\be\label{eqggs11}
\begin{aligned}
{\cal F}_S=&\frac{3 (-64 g^2 (p+1)+8 g (-2 k p+k+2 p+1)-k (4 k p+k-2)-1)}{3p(3 k-1)}+\frac{4 k p + k}{3p(2 \ln(t) + 1)}\\
&-\frac{k (4 p + 3)}{3p(2 \ln(t) + 1)^2}-\frac{8 (1 - 12 g)^2 k ((p + 1) \ln(t) - 1)}{3p(3 k - 
    1) [2 \ln(t) ((6 k - 2) \ln(t) + 3 k - 2) + 3 k - 1]}\\
		&+\frac{4 k}{3p(2 \ln(t) + 1)^3}-\frac{(16 (1 - 12 g)^2 k ((3 k - 2) \ln(t) + 3 k - 1)}{3p(3 k - 
    1) [2 \ln(t) ((6 k - 2) \ln(t) + 3 k - 2) + 3 k - 1]^2}
		\end{aligned}
		\ee
		At large times, taking the limit $t\rightarrow\infty$ in (\ref{eqggs10}) and (\ref{eqggs11}) we find
		\be\label{eqggs12}
		{\cal G}_S|_{t\rightarrow\infty}=\frac{3 (8 g+k-1) (64 g^2+8 g (4 k-1)-6 k^2+3 k-1)}{(1-3 k)^2}
		\ee
		and
			\be\label{eqggs13}
		{\cal F}_S|_{t\rightarrow\infty}=\frac{-1 + 2 k - 64 g^2 (1 + p) - k^2 (1 + 4 p) + 
 8 g (1 + k + 2 p - 2 k p)}{(-1 + 3 k) p}
		\ee
		So the stability under scalar perturbations at large times is possible, provided ${\cal G}_S|_{t\rightarrow\infty}>0$ and ${\cal F}_S|_{t\rightarrow\infty}>0$. Note that the same limits as described by Eqs. (\ref{eqggs12}) and (\ref{eqggs13}) take place at early times $t\rightarrow 0$, and therefore the same conditions guarantee the stability at early and late times.\\
		Taking the limit $p\rightarrow \infty$ in (\ref{eqggs13}), one finds the de Sitter limit 
		\be\label{eqggs14}
		{\cal F}_S|_{dS}=\frac{4(16g^2-4g+4gk+k^2)}{1-3k}
		\ee	
		Note that  ${\cal G}_S|_{t\rightarrow\infty}$ does not depend on $p$. Therefore the solution is stable under scalar perturbations in the de Sitter limit if $g$ and $k$ satisfy the restrictions
		\be\label{eqggs15}
		(8 g+k-1) (64 g^2+8 g (4 k-1)-6 k^2+3 k-1)>0,\;\;\, 16g^2-4g+4gk+k^2>0,\;\;\, 1-3k>0
		\ee
		If we neglect the kinetic coupling ($k=0$), then from (\ref{eqggs10}) and (\ref{eqggs11}) one finds that ${\cal G}_S$ and ${\cal F}_S$ are constants given by
		\be\label{eqggs16}
		{\cal G}_S=3 - 384 g^2 + 1536 g^3,\,\,\,\,\, {\cal F}_S=\frac{1 + 64 g^2 (1 + p) - 8 g (1 + 2 p)}{p}
		\ee
		Thus, in this case the model is free of ghost and gradient instabilities  (independently of time) under scalar perturbations, for the power-law evolution, provided $g$ satisfies the restrictions for $p>0$
		\be\label{eqggs17}
		3 - 384 g^2 + 1536 g^3>0,\,\,\,\, 1 + 64 g^2 (1 + p) - 8 g (1 + 2 p)>0
		\ee
		The accelerated expansion takes place for $p>1$, which together with (\ref{eqggs17}) leads to the restrictions:\\
	\noindent  For $1 < p \le 1 + \sqrt{5}$
		\be\label{eqggs18}
		\frac{1}{16} (1 - \sqrt{5})  < g < \frac{1 + 2 p}{16 (1 + p)} - \frac{1}{16} \sqrt{\frac{4 p^2-3}{(1 + p)^2}},\,\, \text{or},\,\,\, g > \frac{1}{16} (1 + \sqrt{5})
		\ee
		\noindent  For $p > 1 + \sqrt{5}$
		\be\label{eqggs19}
		(1 - \sqrt{5})  < g < \frac{1 + 2 p}{16 (1 + p)} - \frac{1}{16} \sqrt{\frac{4 p^2-3}{(1 + p)^2}},\,\, \text{or},\,\,\, g > \frac{1 + 2 p}{16 (1 + p)} + \frac{1}{16} \sqrt{\frac{4 p^2-3}{(1 + p)^2}}
		\ee
Turning to the second solution (\ref{eq22f}), we considered the case $\eta_2=2$, but the expressions for ${\cal G}_S$ and ${\cal F}_S$ are long enough to be displayed here. We performed the numerical analysis for the same particular case considered for the tensor perturbations. In Fig. 2 we resume the numerical results obtained for the study of stability under scalar perturbations.
\begin{center}
\includegraphics [scale=0.7]{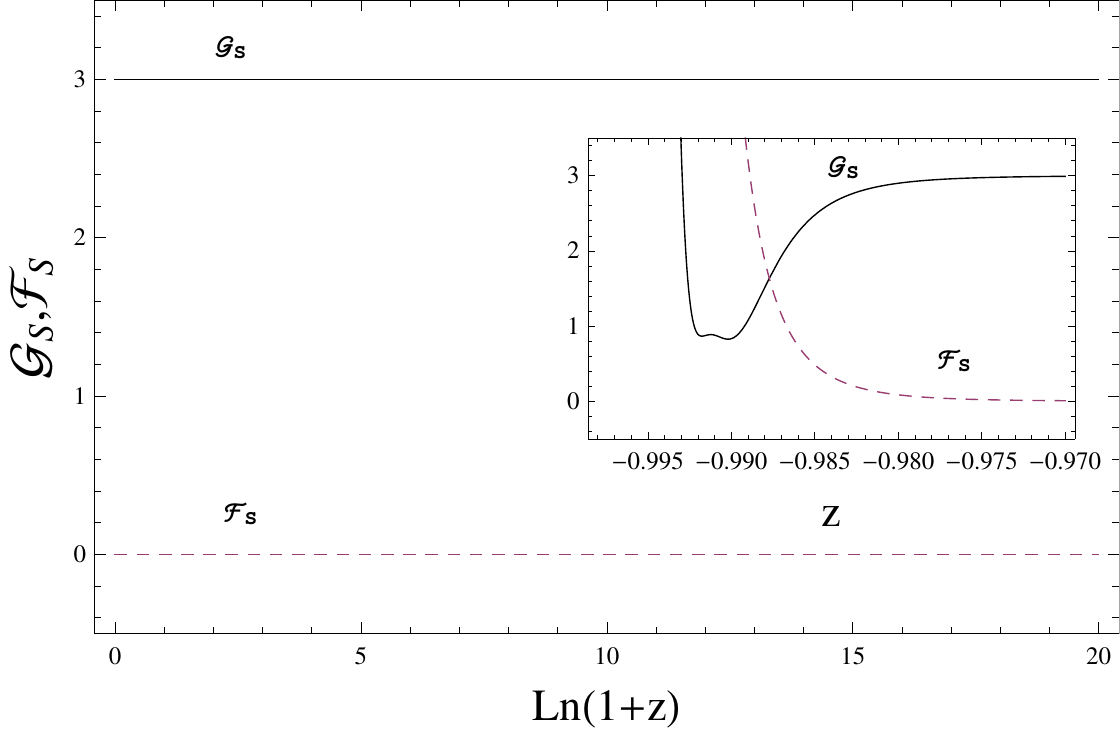}
\end{center} 
\begin{center}
{Fig. 2 \it ${\cal G}_S$ and ${\cal F}_S$ versus $\ln(1+z)$, for $c_g=10^{-12}$ and $c_k=-2\times 10^{-12}$. It was found that ${\cal G}_S=3$ and ${\cal F}_S=0$ with high accuracy during a wide range of redshifts which covers the epochs relevant for dark energy. The internal graphic shows the future behavior of ${\cal G}_T$ and ${\cal F}_T$ for the redshift $z$ in the region $(-0.99,-0.97)$. Note that starting form $z\sim -0.99$ the values of ${\cal G}_S$, ${\cal F}_S$ begin to change}.
\end{center}
Numerical results show that after $z\sim -0.998$ the quantity ${\cal G}_S$ becomes negative and ${\cal F}_S$ starts increasing to large positive values. In summary, we have found that the solution is free of ghost and gradient instabilities under scalar perturbations over a wide range of redshifts, covering different cosmological epochs, including those that are relevant for DE.
\section {Scalar field and matter}
Let's consider the complete model with the contribution of matter sector. Combining the equation of state $p_m=w\rho_m$ with the continuity equation $\dot{\rho_m}+3H(\rho_m+p_m)=0$, leads to the energy density of matter $\rho_m=\rho_{m0}e^{-3(1+w)N}$. It is usually assumed that the matter sector is composed mostly of non relativistic matter with $w=0$. In the e-folding variable $N$, the equation (\ref{eq3}) takes the form
\be\label{eq24a}
H^2=\frac{1}{3}\left[\frac{1}{2}H^2\theta+V+9H^4 F_1\theta+24H^4\frac{dF_2}{dN}+\rho_{m0}e^{-3(1+w)N}\right]
\ee
and the equation of motion (\ref{eq7}) remains as Eq. (\ref{eq22c}). Defining the same limits as before to determine the kinetic and GB couplings, in the e-folding variable take the form
\be\label{eq25}
F_1\theta=\frac{k}{H^2},\,\,\,\,\,\, \frac{dF_2}{dN}=\frac{g}{H^2}
\ee
Using the definition of the density parameters, this corresponds to cosmologically viable limits for the density parameters $\Omega_k=const.=3k$ and $\Omega_{GB}=const.=8g$. 
It remains to determine the potential in order to integrate the equations (\ref{eq24a}) and (\ref{eq22c}). If we demand that $\Omega_{\phi}=const.$, then $\Omega_{DE}=const.$ and independently of the potential the solutions correspond to scaling regime where the energy density of the DE mimics the matter density $\rho_m$. In fact the condition $\Omega_{\phi}=const.$ can be achieved by making $\dot{\phi}^2\propto H^2$ and $V\propto H^2$. But to have a transition to accelerated expansion, we may opt for restrict the potential in such a way that the kinetic term in the Friedmann equation (\ref{eq24a}) behaves like $\dot{\phi}^2\propto H^2$. The restriction 
\be\label{eq26}
V+\rho_{m0}e^{-3(1+w)N}=3\alpha H^2
\ee
reduces the Eq. (\ref{eq24a}) to the form
\be\label{eq27}
\theta=6\left(1-\alpha-3k-8g\right)=\theta_0
\ee
where $1-\alpha-3k-8g>0$. Integrating this equation gives the scalar field as $\phi=\sqrt{\theta_0}N+\phi_0$. The presence of matter term in (\ref{eq26}) marks the difference with scaling behavior. The limit $V\rightarrow 0$ gives rise to scaling solutions, and we should expect that at early times the scalar potential is negligible or is proportional to $\rho_m$. If we introduce the relative potential density $\Omega_V=V/(3H^2)$, then the restriction (\ref{eq26}) is equivalent to $\Omega_V+\Omega_m=\alpha$. Conversely, we may consider (\ref{eq27}) as an ansatz for the free kinetic term (i.e $\dot{\phi}^2/(6H^2)=const.$) (see \cite{neupane}) and arrive at the form of the potential (\ref{eq26}) from the Friedmann equation (\ref{eq24a}). As will be seen the restrictions (\ref{eq25}) and (\ref{eq26}) (or (\ref{eq25}) and (\ref{eq27})) are good approximations for late-time cosmology.
Replacing (\ref{eq25})-(\ref{eq27}) in the equation of motion (\ref{eq22c}) one obtains
\be\label{eq28}
\left(1-k-4g\right)\frac{dH^2}{dN}+2\left(3-3\alpha-6k-20g\right)H^2+(1+w)\rho_{m0} e^{-3(1+w)N}=0
\ee
Satisfaction of one of the two conditions $1-k-4g=0$ or $3-3\alpha-6k-20g=0$, leads to the solution $H^2\propto e^{-3(1+w)N}$ which according to (\ref{eq10}) imply scaling behavior of the DE. Solving (\ref{eq28}) gives
\be\label{eq29}
H^2=H_0^2\left(\eta\Omega_{m0} e^{-3(1+w)N}+\Omega_0 e^{-\gamma N}\right)
\ee
\be\nonumber
\eta=\frac{3}{6\alpha+9k+28g+3w(1-k-4g)-3},\,\,\,\,\, \gamma=2\frac{3\alpha+6k+20g-3}{k+4g-1}
\ee
where $\Omega_0$ is the integration constant and $\Omega_{m0}=\rho_{m0}/(3H_0^2)$ ($H_0$ is the current value of the Hubble parameter). Note that the term $\eta\Omega_{m0}$ can be interpreted as an effective matter parameter. The restriction (\ref{eq10}) imply that $\eta\Omega_{m0}+\Omega_0=1$. From (\ref{eq26}) the scalar potential takes the form
\be\label{eq29a}
V=(\alpha\eta-1)\rho_{m0}e^{-3(1+w)N}+3\alpha H_0^2\Omega_0 e^{-\gamma N}
\ee
Replacing the solution (\ref{eq29}) into Eqs. (\ref{eq25}), one finds the expressions for the kinetic and GB couplings as
\be\label{eq29a1}
F_1=\frac{k}{\theta_0 H_0^2}\left(\eta\Omega_{m0} e^{-3(1+w)N}+\Omega_0 e^{-\gamma N}\right)^{-1}
\ee
where $\theta_0$ is given by (\ref{eq27}). 
\be\label{eq29a2}
F_2=\frac{g e^{\gamma N}}{\gamma\Omega_0 H_0^2}\hspace{0.2cm} _{2}F_1\left[-\frac{\gamma}{3w-\gamma+3},1,\frac{3w-2\gamma+3}{3w-\gamma+3},-\frac{\eta\Omega_{m0} e^{-(3w-\gamma+3)N}}{\Omega_0}\right]
\ee
An interesting case takes place in Eq. (\ref{eq29a}) when $\alpha\eta=1$. In this case the matter term in the potential disappears, giving place to the simple exponential behavior
\be\label{eq29b}
V=3\alpha H_0^2\Omega_0 e^{-\gamma N}=3\alpha H_0^2\Omega_0 e^{-\gamma \phi/\sqrt{\theta_0}}
\ee
where $-3\alpha=9k+28g+3w(1-k-4g)-3$, and we used the solution for the scalar field with $\phi_0=0$.
According to the solution (\ref{eq29}) and Friedmann equation (\ref{eq3}) the DE density is given by
\be\label{eq30}
\rho_{DE}=(\eta-1)\rho_{m0}e^{-3(1+w)N}+3\Omega_0 H_0^2e^{-\gamma N}=(\eta-1)\rho_{m0}(1+z)^{3(1+w)}+3\Omega_0 H_0^2(1+z)^{\gamma}
\ee
where we used the redshift relation: $N=-\ln(1+z)$. It is clear that the scalar field keeps track of the DM at early times if $\gamma<3(1+w)$  (or more sharply if $\gamma<0$, since in this case the second term becomes negligible for $z>>1$). This solution agree with the fact that during the matter dominated epoch $w_{DE}\sim w$. The first term in (\ref{eq29}) is an adjustment of the initial parameter appearing in the Lagrangian density of the matter due to the interacting terms, which we interpret as an effective matter parameter. Therefore, is the quantity $\eta\Omega_{m0}$ which has to be adjusted with the observational data and should be about $0.27$.
If $\eta=1$, the DE presents power-law behavior with constant EoS $w_{DE}=-1+\gamma/3$. The condition $\gamma<0$ in (\ref{eq30}) leads to quintom behavior, since it allows the transition to the phantom phase.\\
In the important case of non relativistic dark matter component with $w=0$, the DE equation of state that follows from (\ref{eq30}) is 
\be\label{eq31}
w_{DE}=-\frac{(1-\eta\Omega_{m0})(3-\gamma)e^{-\gamma N}}{3(1-\eta\Omega_{m0})e^{-\gamma N}+3(\eta-1)\Omega_{m0}e^{-3N}}
\ee
the current value $w_{DE}(0)=w_0$ may be written as
\be\label{eq32}
w_0=\left(\frac{1-\eta\Omega_{m0}}{1-\Omega_{m0}}\right)\left(-1+\frac{\gamma}{3}\right)
\ee
The first factor in this equation qualifies the difference with respect to constant EoS, which is the case for $\eta=1$. According to current observations $w_{DE}$ could be very close to $-1$; so setting $w_0=-1$ in (\ref{eq32}) we obtain the relation
\be\label{eq33}
\gamma=\frac{3(1-\eta)\Omega_{m0}}{1-\eta\Omega_{m0}}
\ee
On the other hand, the deceleration-acceleration transition occurs at redshift $z_T$ when $w_{eff}=-1/3$. Evaluating $w_{eff}=-1-\frac{2\dot{H}}{3H^2}$ from (\ref{eq29}), and setting $w_{eff}(z_T)=-1/3$ we find another relationship
\be\label{eq34}
\eta=\frac{3(2-\gamma)}{3(2-\gamma)+\gamma(1+z_T)^{3-\gamma}}
\ee
where we used (\ref{eq33}). Thus for example, taking $z_T=0.7$ and $\gamma=-0.1$, gives $\eta\approx 1.09$ and from (\ref{eq33}), $\Omega_{m0}\approx 0.26$. So, the restrictions (\ref{eq25}) and (\ref{eq26}) allow to describe a viable cosmological scenario.\\
We can make another interesting choice for the kinetic and GB couplings, namely we consider the ansatzes for the density parameters 
\be\label{eq35}
\Omega_k=c_k e^{-\lambda N},\,\,\,\,\, \Omega_{GB}=c_g e^{-\lambda N}
\ee
this means that $\Omega_k, \Omega_{GB}$ scale as $a^{-\lambda}$, where $\lambda$ should take such value that guarantees the early time dominance of the matter sector and lower rate of decaying of dark energy during expansion at late times. With this choice the Friedmann equation (\ref{eq24a}) becomes
\be\label{eq36}
3H^2=\frac{1}{2}H^2\theta+V+3(c_k+c_g)H^2e^{-\lambda N}+\rho_{m0}e^{-3(1+w)N}
\ee
and restricting the potential in the form
\be\label{eq37}
V+3(c_k+c_g)H^2e^{-\lambda N}+\rho_{m0}e^{-3(1+w)N}=3\alpha H^2,
\ee
reduces the Eq. (\ref{eq36}) to: $\theta=\theta_0=6(1-\alpha)$, where $0\le\alpha<1$. Replacing (\ref{eq35}), (\ref{eq37}) and $\theta$ in the equation of motion (\ref{eq22c}) one obtains
\be\label{eq38}
\begin{aligned}
\left(3-(c_k+3c_g/2)e^{-\lambda N}\right)\frac{dH^2}{dN}&+\left(18(1-\alpha)+\left[2(3+\lambda)c_k+3(1+\lambda)c_g\right]e^{-\lambda N}\right)H^2\\
&+3(1+w)\rho_{m0}e^{-3(1+w) N}=0
\end{aligned}
\ee
In order to simplify the integration of this equation we take $\alpha=\frac{24(c_g+c_k)+3c_g\lambda+2c_k\lambda}{6(3c_g+2c_k)}$, and considering the important case of non relativistic presureless matter with $w=0$,  we find
\be\label{eq39}
H^2=\frac{6\sigma_1\rho_{m0}(1+z)^3}{(6 -\sigma_1(1+z)^{\lambda})\sigma_2}+\frac{C(1+z)^{-2\sigma_3/\sigma_1+\lambda}}{6-\sigma_1(1+z)^{\lambda}}
\ee
where we used the relation: $N=-\ln(1+z)$, and $\sigma_1=3c_g+2c_k$, $\sigma_2=3c_g(\lambda+5)+2c_k(\lambda+9)$ and $\sigma_3=3c_g(\lambda+1)+2c_k(\lambda+3)$. Using this result and 
Eqs. (\ref{eq25}) we find the following expressions for the kinetic and GB couplings
\be\label{eq39a}
F_1=\left(\frac{k}{\theta_0}\right)\frac{6-\sigma_1 e^{-\lambda N}}{\frac{6\sigma_1}{\sigma_2}\rho_{m0}e^{-3N}+Ce^{(2\sigma_3/\sigma_1-\lambda )N}},
\ee
where $\theta_0=6(1-\alpha)$, and
\be\label{eq39b}
F_2=\frac{6\rho_{m0}e^{-(3-\lambda)N}}{(3-\lambda)\sigma_2}\hspace{0.2cm} _{2}F_1\left[1,\frac{\lambda-3}{\lambda},2-\frac{3}{\lambda},\frac{6e^{\lambda N}}{\sigma_1}\right]
-\frac{Ce^{2\sigma_3 N/\sigma_1}}{2\sigma_3}\hspace{0.1cm} _{2}F_1\left[1,\frac{2\sigma_3}{\lambda\sigma_1},1+\frac{2\sigma_3}{\lambda\sigma_1},\frac{6e^{\lambda N}}{\sigma_1}\right]
\ee
Note that the solution (\ref{eq39}) has sense only if $\lambda<0$. In this case at early times ($z\rightarrow\infty$), the Hubble parameter behaves as $H^2\sim (\sigma_1/\sigma_2)\rho_{m_0}(1+z)^3+(C/6)(1+z)^{\lambda-2\sigma_3/\sigma_2}$. The first term is the usual matter term and the second term behaves as $(1+z)^{\lambda-2\sigma_3/\sigma_2}$. So to maintain matter dominance at early times is sufficient the condition $\lambda-2\sigma_3/\sigma_2<2$, which satisfies the nucleosintesis constraint (with appropriate constant $C$) on dark energy in early cosmology. An interesting fact of the solution (\ref{eq39}) is the existence of future singularity at the redshift $z_s=(6/\sigma_1)^{1/\lambda}-1$, where $\sigma_1$ should satisfy the inequality $0<(6/\sigma_1)^{1/\lambda}<1$ in order to have the singularity in the future. Is easy to see from (\ref{eq39}) that at this point $\dot{H}\rightarrow\infty$, which together with the singularity in the Hubble parameter signify that the energy density and pressure become infinity at $z_s$. By definition, at this point $a=1/(z_s+1)$ is finite, so we have a type III singularity (\cite{sergeio5}, \cite{sergeio6}). Note also that from Eqs. (\ref{eq39a}) and (\ref{eq39b}), it follows that the kinetic coupling disappears ($F_1\rightarrow 0$) and the GB coupling diverges at the singular point $z_s$. From Eq. (\ref{eq39}), the DE equation of state is given by (assuming presureless matter)
\be\label{eq40}
w_{DE}=\frac{(1+z)\frac{d\tilde{H}^2}{dz}-3\tilde{H}^2}{3\tilde{H}^2-3\Omega_{m0}(1+z)^3}
\ee
where $\tilde{H}=H/H_0$ and $H_0$ is the current value of the Hubble parameter. As an example, a set of parameters satisfying all discussed above restrictions is: $\sigma_1=4, \sigma_2=18, \sigma_3=1, \lambda=-1/4, \Omega_{m0}=0.27, \tilde{C}=0.92$, and the constant $C$ is related to the other parameters through the flatness condition: $\tilde{C}=C/H_0^2=6-\sigma_1-18\sigma_1\Omega_{m0}/\sigma_2$. For this set of parameters,  at early times $w_{DE}|_{z\rightarrow\infty}=0$, which means that at early times the DE mimics the usual matter behavior (or early time matter dominance as follows from (\ref{eq39}). The current value of DE EoS is $w_{DE}(0)\approx -1.02$, and the asymptotic value at far future is $w_{DE}|_{z\rightarrow\infty}=-1-1/6$. The type III future singularity takes place at $z_s\approx -0.8$. Note that although the type III singularities are weaker than big rip singularities, and in some cases one can extend the evolution behind this singularity \cite{cotsakis}, \cite{cotsakis1}, \cite{dabrowski}, in the concrete case considered here, although the density becomes finite after the singularity, it also becomes negative (i.e $H^2<0$ for $t>t_s$), which makes it impossible for the universe to evolve beyond this singularity. Therefore, one can distinguish an scenario of the singularity of type III, which is stronger in the sense that the evolution beyond it, is not physically allowable.
\section{Discussion}
We studied late time cosmological solutions based on string spired scalar-tensor model including a coupling to the Gauss-Bonnet invariant and kinetic couplings to curvature. We presented various exact cosmological solutions that exhibit the deceleration-acceleration phase and even the very probably phantom phase. These phases are 
necessary for the successful explanation of the early inflation as well as the currently observed phase of acceleration (super-acceleration) in late universe.\\ 
As a criteria to fix the couplings we have based on asymptotic limits of the model. Thus in absence of potential, the limit of vanishing time derivative of the scalar field ($\dot{\phi}\sim 0$) suggests the constraint (\ref{eq13}) imposed on the GB coupling. Neglecting the kinetic term, the limit ($F_2\rightarrow 0$, $V\rightarrow 0$) suggests a cosmologically viable restriction on the kinetic coupling (\ref{eq9}). The limit ($F_1\rightarrow 0, F_2\rightarrow 0$,  $V\rightarrow 0$) suggests a behavior of the kinetic term (\ref{eq15}) which is sufficiently general to obtain interesting cosmological scenarios. The restrictions on the interaction terms are also equivalent to the constancy of the respective density parameters, i.e. $\Omega_k=const.$ and $\Omega_{GB}=const.$ (see (\ref{eq13a})), that are reasonable limits for a late time universe. These restriction in the scalar field dominance case, lead to the exponential couplings characteristic of the low energy $\alpha'$ expansion in string theory \cite{tseytlin}, \cite{cartier}, \cite{meissner}. 
In the more realistic case with the presence of matter, the restrictions (\ref{eq13a}) (or (\ref{eq25})) lead to solution (\ref{eq29}) with dark energy that keeps track of the earlier matter dominance, and containing late time quintessence and phantom phases. In this case, to maintain the behavior of the kinetic  term as $\dot{\phi}\propto H^2$, we restrict the scalar potential according to the expression (\ref{eq26}), which leads to the exponential behavior (\ref{eq29a}, \ref{eq29b}).\\ 
Another cosmologically viable expressions for the kinetic and GB couplings are obtained by imposing scaling behavior of the type $a^{-\lambda}$ to the corresponding density parameters, i.e. $\Omega_k\propto a^{-\lambda}$ and $\Omega_{DE}\propto a^{-\lambda}$, where we considered the same power in order find exact analytical solutions. In the case of scalar field dominance, we found a cosmological solution (\ref{eq22f}) able to explain late time accelerated expansion and even a transition to phantom phase. We considered the autonomous system for the model in absence of matter and have found the conditions for stability of the solutions (\ref{eq19}) and (\ref{eq22f}). We have also studied the stability properties of the solutions (\ref{eq19}) and (\ref{eq22f}) under tensor and scalar perturbations of the metric. For the power-law solution we have found the restrictions on the parameters in order to avoid ghost and gradient instabilities in the following cases: at large times, in the de Sitter limit, and in absence of the kinetic coupling. For the solution (\ref{eq22f}) we considered a DE scenario very close (with high accuracy) to the de Sitter solution, and it was proved that the solution is free of ghost and gradient instabilities under tensor and scalar perturbations, from early times to the present and into the future up to $z\sim - 0.99$.\\
In the model with matter content, a variety of exact solutions can be found depending on the correlation between the parameters of the equation (\ref{eq38}). In fact, the general solution to the cosmological equation (\ref{eq38}) is given through hyper-geometric function, but using the limits when this function simplifies to elementary exponentials we found relations between the parameters. For one particular choice of the parameters, we found the solution (\ref{eq39}), which unifies early time matter dominance with late time accelerated expansion including transition to phantom phase. This solution also presents future type III singularity provided $\sigma_1$ in (\ref{eq39}) satisfies $0<(6/\sigma_1)^{1/\lambda}<1$. An specific characteristic of the singularity that appears in the solution (\ref{eq39}) is that the density becomes negative for times $t>t_s$ making it impossible for the universe to evolve beyond this singularity.
We cited one example, but there are another relations between the parameters of Eq. (\ref{eq38}) that could realize phenomenologically acceptable cosmological scenarios. 
In all considered solutions the scaling behavior of the kinetic energy is the same as of the total energy density, i.e. $\dot{\phi}^2\propto H^2$, which in absence of interactions corresponds to a conformal limit of the model (a purely free kinetic scalar model is conformal invariant). 
We have studied the dark energy problem in the context of the scalar-tensor model, with kinetic and GB couplings contributing to the total energy density. By imposing physically motivated restrictions on density parameters involving the interaction terms, we significantly simplified the cosmological equations and have obtained exact viable solutions, compatible with the current phenomenology of the dark energy. 

\end{document}